\documentclass{ifacconf}

\usepackage{graphicx}      
\usepackage{natbib}        
\usepackage{color}   
\usepackage{amsmath}
\usepackage{amssymb}
\usepackage{algorithmicx}
\usepackage{algorithm}
\usepackage{algpseudocode}
\usepackage{stfloats}
\usepackage{calc}  
\usepackage{enumitem}  
\usepackage{url}
\algdef{SE}[DOWHILE]{Do}{doWhile}{\algorithmicdo}[1]{\algorithmicwhile\ #1}%
\begin{document}
\begin{frontmatter}

\title{Initialization of a Disease Transmission Model\thanksref{footnoteinfo}} 

\thanks[footnoteinfo]{This work is funded in part by the PhD program at the Centre for Interdisciplinary Mathematics, Uppsala University, Sweden, and by the Swedish Research Council, under the grant 2019-04451.}

\author[First]{H\aa kan Runvik} 
\author[First]{Alexander Medvedev} 
\author[First]{Robin Eriksson}
\author[First]{Stefan Engblom}

\address[First]{ Information Technology, Uppsala University, Uppsala, SWEDEN, e-mail: \texttt{\{hakan.runvik, alexander.medvedev, robin.eriksson, stefan.engblom\}@it.uu.se}.}

\begin{abstract}                
Approaches to the calculation of the full state vector of a lager epidemiological model for the spread of COVID-19 in Sweden at the initial time instant from available data and with a simplified dynamical model are proposed and evaluated. The larger epidemiological model is based on a continuous Markov chain and captures the demographic composition of and the transport flows between the counties of Sweden. Its intended use is to predict the outbreak development in temporal and spatial coordinates as well as across the demographic groups. It can also support evaluating and comparing of prospective intervention strategies in terms of e.g. lockdown in certain areas or isolation of specific age groups. The simplified model is a discrete time-invariant linear system that has cumulative infectious incidence, infected population, asymptomatic population, exposed population, and infectious pressure as the state variables. Since the system matrix of the model depends on a number transition rates, structural properties of the model are investigated for suitable parameter ranges. It is concluded that the model becomes unobservable for some parameter values. Two contrasting approaches to the initial state estimation are considered. One is a version of Rauch–Tung–Striebel smoother and another is based on solving a batch nonlinear optimization problem. The benefits and shortcomings of the considered estimation techniques are analyzed and compared on synthetic data for several Swedish counties.
\end{abstract}


\end{frontmatter}

\section{Introduction}

\begin{figure}[ht]
    \centering
    \includegraphics[width=8.0cm]{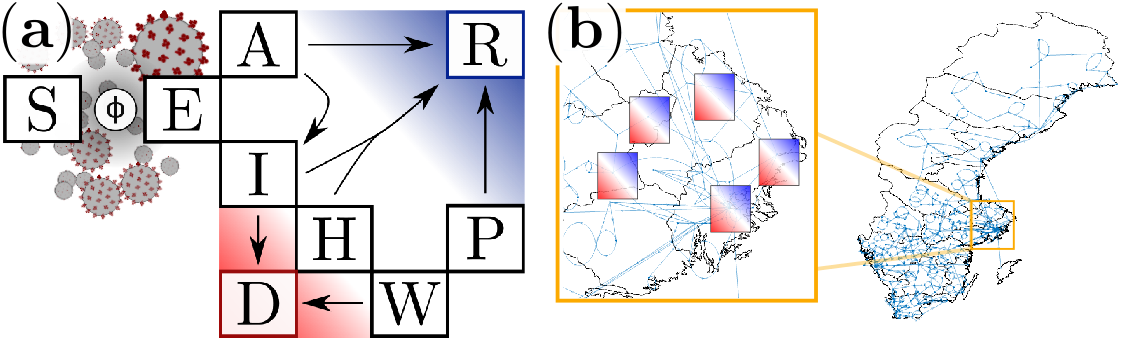}
    \caption{\textbf{(a)} Illustration of the compartment model of the Swedish COVID-19 outbreak. Arrows denote the flows of individuals between the compartments: susceptible (S), exposed (E), asymptomatic (A), symptomatic (I), hospitalized (H), intensive care (W), post-intensive care (P), deceased (D), recovered (R). \textbf{(b)} The full model is a network of compartment models emulating the commuting network between the Swedish municipalities.
    The yellow rectangle shows a zoom-in view of the network within the greater Stockholm region.}\label{fig:cov19}
\end{figure}

This paper is concerned with using publicly available epidemiological data for estimating suitable initial conditions for a large mechanistic general Susceptible-Exposed-Infectious-Recovered (SEIR) model of the Swedish COVID-19 outbreak. The model incorporates spatial communication between the Swedish municipalities, and also includes the Swedish demographics, thought to be an important factor for the impact of COVID-19 \cite{keeling2008modeling}. The viral contraction is driven by an infectious pressure as in \cite{siminf3, engblom2019bayesian}. Fig.~\ref{fig:cov19} provides an overview of the modeling approach and specifies the included compartments.

The dynamics of the disease transmission are modeled by a discrete-state continuous-time Markov chain. A continuous state variable, the environmental compartment, is included to model the infectious pressure. The  Markov chain model is implemented using the computational framework SimInf in R, \cite{siminf_manual}. To infer the model parameters, the aim is to utilize a Bayesian approach as it allows the use of empirical measures as prior knowledge of the model parameters.

The problem of  estimating the state vector of a dynamical system backwards in time is known as smoothing.  An optimal (minimal variance) fixed-interval smoother for a linear time-invariant model under additive Gaussian noise assumption was derived in \cite{RTS65}. Since then, various methods have  been devised for more general settings, including state-dependent Gaussian noise (\cite{ab12}) and non-Gaussian noise sources (\cite{wzw20}). 
In the present work, these two complications occur combined, as the process noise is Poisson-distributed rather than Gaussian, and also dependent on the plant state. Therefore, none of the approaches found in the literature is readily applicable here. Instead, to obtain a plausible solution fast, empirical initialization algorithms are developed and compared to determine which one is most suitable in the final setup. To establish ground truth, synthetic data produced by models of increasing complexity are utilized in the performance evaluation.

The rest of the paper is organized as follows. First, the model initialization problem is formulated and the properties of the linear time-invariant model that is used to calculate the initial condition are explored. Then, three model-based approaches to solving the initialization problem are presented. Finally, performance of the considered approaches is evaluated on synthetic data and conclusions are drawn.

\section{Model initialization problem}

The inputs to the Markov chain model are the parameters inferred from data and an initial chain state. The initial state consists of the epidemiological states in all compartments, including the hidden states, i.e., the exposed and asymptomatic carriers.

To find a county-wise initialization, specific to the Swedish COVID-19 outbreak, the cumulative infected cases data reported by the Swedish public health agency were employed \cite{FHM_data}. In Sweden, a full disease testing strategy was in effect until March 12, after which the testing was heavily restricted \cite{FHM_testing}. With full testing, we assume that the reported cases holds the true number of cumulative infected cases.

An accepted standard in stochastic epidemiological modeling is to start simulations when the system has reached some (fairly large) threshold number \cite{allen2017primer, giordano2020modelling}. We used the threshold of a 100 reported cases which Sweden reached on March 6; the data up until March 12 can therefore be used for smoothing.

The problem of estimating the infected, exposed, and asymptomatic populations at a given point in time (model initialization point) is therefore investigated, based on the data for  cumulative incidence measured over a fixed time horizon. Thus, the problem at hand  constitutes  a fixed-interval smoothing problem. The remaining compartments of the Markov chain model do not influence the infected, exposed or asymptomatic populations and are therefore not included at present in the considered estimation problem.

\subsection{Initialization data}

Epidemiological mathematical models are typically designed in terms of populations and face difficulties in capturing situations, when only a few individuals are infected. This is logically the case in the beginning of an outbreak. Besides, an epidemic is not readily recognized until the number of patients in the healthcare system becomes significant, thus making initial data scarce and unreliable. Yet, since disease transmission is a dynamical process, a mathematical model of it has to be  initialized so that historical data for the observed output agree well with the  output produced by the model.

As there were no deaths from the disease and very few individuals were in intensive care prior to the chosen point of initialization, the measurements that are used as input to the Markov chain model cannot be used for the initialization of it. Instead, reported county-wise cumulative incidence from the period of February 4th to March 12th 2020 are utilized. As contact tracing was discontinued after this period, incidence data from later times are significantly less reliable.

\subsection{Initialization model}

Since direct inversion of a continuous Markov chain is not easily apprehended, the following linear time-invariant approximation is utilized for the initialization of the model for each county, whereas the model states are lumped over the considered age groups. The latter simplification is introduced since the cases were few in the beginning of the outbreak and patient age was not specified in the data.

The model is derived as a normal approximation of the Poisson distributed forward steps and formulated in state-space form as
\begin{equation}
    x_{k+1}=F x_k + w_k, \label{eq:LTI}
\end{equation}
where
\begin{equation*}
    F=\begin{bmatrix} 1 & 0 & \gamma_\mathrm{A} F_1 & \sigma F_0 & 0 \\
    0 & 1-\gamma_\mathrm{I} & \gamma_\mathrm{A} F_1 & \sigma F_0 & 0 \\
    0 & 0 & 1-\gamma_\mathrm{A} & \sigma (1-F_0) & 0 \\
    0 & 0 & 0 & 1-\sigma & \beta \\
    0 & 1-\e^{-\rho} & \theta_\mathrm{A} (1-\e^{-\rho}) & \theta_\mathrm{E}(1-\e^{-\rho}) & \e^{-\rho}
    \end{bmatrix},
\end{equation*}
$k=0,1,\dots$ is the discrete time corresponding to daily sampling and $w_k$ is the process noise sequence, whose properties will be clarified in Section \ref{sec:noiseCov}.
The state vector elements
\[x_k=\begin{bmatrix}I_{c}(k) & I(k) & A(k) & E(k) & \phi(k)  \end{bmatrix}^\intercal\]
stand for the populations of the model compartments according to: 
\begin{description}[leftmargin=!,labelwidth=\widthof{\bfseries AAA}]
    \item[$I_c$] cumulative infectious incidence,
    \item[$I$] infected,
    \item[$A$] asymptomatic,
    \item[$E$] exposed,
    \item[$\phi$] infectious pressure.
\end{description}

The parameters of the model are specified below
\begin{description}[leftmargin=!,labelwidth=\widthof{\bfseries AAA}]
    \item[$\sigma$] expected rate of transition from the exposed state,
    \item[$\gamma_\mathrm{A}$] expected rate of transition from asymptomatic state,
    \item[$\gamma_\mathrm{I}$] expected rate of transition from infected state,
    \item[$F_0$] fraction of transition from exposed reaching the infected state; the remaining fraction reaches the asymptomatic state,
    \item[$F_1$] fraction of transition from asymptomatic state reaching the infected state, The remaining fraction corresponds to the recovery from the disease (not included in \eqref{eq:LTI}),
    \item[$\beta$] indirect transmission rate of the environmental infectious pressure,
    \item[$\rho$] infections pressure decay rate,
    \item[$\theta_\mathrm{A}$] asymptomatic viral shedding  rate,
    \item[$\theta_\mathrm{E}$] exposed viral shedding  rate. 
\end{description}
The parameters are positive and so are the elements of the state matrix $F$. Therefore, model \eqref{eq:LTI} is also positive, i.e. the state vector belongs to the positive quadrant provided the initial condition $x_0$  and $w_k, k=0,1,\dots$ do. The latter condition restricts the distribution of the process noise.

To obtain the parameter values for model \eqref{eq:LTI}, prior distributions for the Bayesian parameter estimation algorithm of the Markov chain model are utilized. The prior distributions are based on empirical data or published estimates. 

For parameter values from these distributions, the matrix $F$ tends to have one eigenvalue with magnitude larger than one and is therefore unstable. This is expected, since exponential growth is observed during the early phase of a disease outbreak. 

Since the cumulative incidence is the only measured signal, the output of the model is
\begin{equation}\label{eq:LTIout}
    y_k = H x_k + v_k,
\end{equation}
where
\begin{equation*}
    H = \begin{bmatrix}1 & 0 & 0 & 0 & 0\end{bmatrix},
\end{equation*}
and $v_k$ is the measurement noise with zero mean and variance $R_k$. The introduction of measurement noise is a matter of complying with the standard assumptions of Kalman filtering and not an actual model property.

 Model \eqref{eq:LTI}, \eqref{eq:LTIout}  does not possess structural observability for the whole range the parameter values. Some combinations of parameter values sampled from the prior distribution make the observability matrix
\begin{equation*}
    \mathcal{O}= \begin{bmatrix}H \\ HF \\ HF^2 \\ HF^3 \\HF^4 \end{bmatrix}.
\end{equation*}
lose rank. 

\begin{figure*}[b]
\begin{equation}\label{eq:covQ}
    {Q_1}_k = \begin{bmatrix}
    \gamma_\mathrm{A} F_1 A(k) +\sigma F_0 E(k) & \gamma_\mathrm{A} F_1 A(k) +\sigma F_0 E(k) & -\gamma_\mathrm{A} F_1 A(k) & -\sigma F_0 E(k) & 0 \\ \gamma_\mathrm{A} F_1 A(k) +\sigma F_0 E(k) & \gamma_\mathrm{A} F_1 A(k) +\sigma F_0 E(k)+ \gamma_\mathrm{I} I(k) &  -\gamma_\mathrm{A} F_1 A(k) & -\sigma F_0 E(k) & 0 \\
    -\gamma_\mathrm{A} F_1 A(k) & -\gamma_\mathrm{A} F_1 A(k) & \gamma_\mathrm{A} A(k) + \sigma (1-F_0) E(k) & -\sigma (1-F_0) E(k) & 0\\ -\sigma F_0 E(k) & -\sigma F_0 E(k) & -\sigma (1-F_0) E(k) & \sigma E+\beta \phi(k) & 0 \\ 0 & 0 & 0 & 0 & 0
    \end{bmatrix}
\end{equation}
\end{figure*}

\subsection{Process noise covariance}
\label{sec:noiseCov}

In order to analyze the process noise covariance, each error vector $w_k$ is separated into two terms:
\begin{equation*}
    w_k = {w_{1}}_k + {w_0}_k,
\end{equation*}
where ${w_{1}}_k$ describes the error of approximating the stochasticity of the full Markov chain model by the linear dynamics of \eqref{eq:LTI}, and ${w_0}_k$ captures any other model uncertainty, including both differences between the models (e.g. the spread between counties) and differences between the complete model and the true outbreak dynamics. The process noise covariance matrix $Q_k$ is  split accordingly as
\begin{equation*}
    Q_k={Q_1}_k+Q_0.
\end{equation*}
The model uncertainty is assumed to be additive, independent of $k$, and uncorrelated between the components.  Therefore, $Q_0$ is diagonal and constant.

The evaluation of the approximation error covariance ${Q_1}_k$ is more challenging. When the Markov chain model is sampled, the distributions of the elements of ${w_{1}}_k$ are given by sums of Poisson processes that are shifted to have zero mean, and with variance that depend on the populations in the different compartments.

The matrix ${Q_1}_k$ is thus state-dependent and  evaluated  to \eqref{eq:covQ}. To avoid confusion with pure time-varying case, the explicit notation is utilized
\begin{equation*}
    Q_k=Q(x_k).
\end{equation*}

\section{Smoothing problem}

Let $I_\mathrm D =[0,d]$ define a finite interval of discrete time instants corresponding to the measurements  $y_k, k\in I_\mathrm D$, and  $m\in I_\mathrm D$ be the point of initialization of the Markov chain model. 

An estimate $\hat{x}_{m|d}$ of $x_k|_{k=m}$ defined by model \eqref{eq:LTI} is then sought  from   the output data $y_k, k\in I_\mathrm D$.
The  problem  at hand was approached using three different methods, which are presented next.

\subsection{Rauch-Tung-Striebel smoother}
The Rauch-Tung-Striebel (RTS) smoother (\cite{RTS65}) is a recursive method for solving fixed-interval smoothing problems. It is proven to be an optimal smoother,  when the noise sources are Gaussian and independent of the system states, and lacks theoretical justification in the present case. Even stability properties of the RTS smoother are not readily guaranteed.
However, as the results of  Section~\ref{sec:experiments} demonstrate, it can nonetheless be used empirically. The stability concerns are not critical as the estimation is performed with a discrete LTI model and on a finite time interval.

The RTS smoother is a two-pass algorithm consisting of a Kalman filter that is run for the full interval in a forward pass, followed by a backwards pass, when the state estimates are smoothed. The Kalman filter equations that are solved recursively from the initial conditions $\hat x_0$ and $P_{0|0}$ are
\begin{align*}
    \hat{x}_{k|k-1}&=F\hat{x}_{k-1|k-1}, \\
    P_{k|k-1}&=F P_{k-1|k-1} F^{\intercal} + Q_k,\\
    \tilde{y}_k&=y_k-H\hat{x}_{k|k-1},\\
    S_k&=H P_{k|k-1} H^{\intercal} + R_k,\\
    K_k&=P_{k|k-1} H^{\intercal} S_k^{-1},\\
    \hat{x}_{k|k}&=\hat{x}_{k|k-1} + K_k \tilde{y}_k\\
    P_{k|k}&=(I-K_k H) P_{k|k-1},\\
\end{align*}
where $\hat{x}_{k|k-1}$ and $\hat{x}_{k|k}$ are the {\it a priori} and  {\it a posteriori} state estimates, $P_{k|k-1}$ and $P_{k|k}$ are the {\it a priori} and {\it a posteriori} estimate covariances, $\tilde{y}_k$ is the innovation, $S_k$ is the innovation covariance, and $K_k$ is the Kalman gain.

Notice that the Kalman filter requires knowledge of the covariance matrix $Q_k$ for $1\le k\le d$. In the present case, the covariance matrix is not available, since it depends on the unknown states of the system. Therefore, the  plant state is replaced by its estimate, and the covariance matrix $Q_k$ is approximated as 
\begin{equation*}
    \hat Q_k=Q(\hat{x}_{k|k-1}).
\end{equation*}

The {\it a priori} and {\it a posteriori} state and covariance estimates at each time are are saved for the backwards pass. Then the algorithm proceeds backwards from the last time point $d$. The smoothed estimate $\hat{x}_{k|d}$ is calculated recursively via the equations
\begin{align*}
    \hat{x}_{k|d} &= \hat{x}_{k|k}+C_k(\hat{x}_{k+1|d}- \hat{x}_{k+1|k}), \\
    P_{k|d}&=P_{k|k}+C_k(P_{k+1|d}+P_{k+1|k})C_k^\intercal,
\end{align*}
where $C_k=P_{k|k} F^{\intercal} P_{k+1|k}^{-1}$ and  $P_{k|d}$ is the smoothed estimate covariance.  

\subsection{Ordinary least squares}
The  problem of estimating $\hat{x}_{m|d}$ can be approached as an optimization problem and solved once, rather than recursively. The simplest setup is based on the linear relation between the measurement and the state (i.e. backcasting) and leads to the algebraic system
\begin{equation*}
    y_d=H F^{k-d} x_k + \tilde{w}_k,
\end{equation*}
where the properties of the noise $\tilde{w}_k$ will be elaborated upon in Section \ref{sec:NLS}. The state estimation problem is then formulated as \begin{equation}\label{eq:OLS}
    \hat{x}_{m|d} = \arg \min_{x_m} ||Y-\Phi x_m||^2,
\end{equation}
where
\begin{equation*}
    Y=\begin{bmatrix}y_1 & y_2 & \dots & y_d \end{bmatrix}^\intercal,
\end{equation*}

\begin{equation*}
    \Phi=\begin{bmatrix}
        HF^{1-m} \\ HF^{2-m} \\ \vdots \\ HF^{d-m}
    \end{bmatrix}.
\end{equation*}

Optimization problem \eqref{eq:OLS} can be solved using standard techniques for linear least squares. Furthermore, positivity of the state estimation can be enforced by using constrained least squares.

\subsection{Nonlinear least squares}
\label{sec:NLS}

The basic method presented above can be potentially  improved through weighting  by taking into account the correlation of the error terms $\tilde{w}_k$. To this end, let $S_\mathrm Q =\{Q_k\}_{k=0}^d$, and define the matrix $\Omega(S_\mathrm Q)$ by specifying its elements as
\begin{equation*}
    \omega_{k,l}=\sum_{n=0}^{r_1-1} H F^{r_2} Q_{r_3} {F^\intercal}^{\tilde{n}} H^\intercal+\delta_{kl} R_k,
\end{equation*}
where
\begin{equation*}
\tilde{n}=\begin{cases}
n, & k<m \\ -n-1, & k \ge m
\end{cases},
\end{equation*}
\begin{equation*}
    r_1=\begin{cases} \min(|k-m|,|l-m|), & (k-m)(l-m) \ge 0 \\ 0, & (k-m)(l-m)<0
    \end{cases},
\end{equation*}
\begin{equation*}
    r_2=\begin{cases} \tilde{n}-k+l, & |k-m|\le|l-m| \\
    \tilde{n}-l+k, & |k-m|>|l-m|
    \end{cases},
\end{equation*}

\begin{equation*}
    r_3=\begin{cases} k-1-\tilde{n}, & |k-m|\le|l-m| \\
    l-1-\tilde{n}, & |k-m|>|l-m|
    \end{cases}.
\end{equation*}
Then, $\Omega(S_\mathrm Q)$ is the covariance matrix of 
$$W = \begin{bmatrix}\tilde w_0 & \tilde  w_1 & \dots & \tilde  w_d \end{bmatrix}^\intercal.$$
The error terms $\tilde{w}_k$ are thus neither uncorrelated nor homoscedastic, so the Gauss-Markov theorem does not apply to the ordinary least squares formulation in \eqref{eq:OLS}.

If the process noise covariance matrices were independent of the system states, the best linear unbiased estimator would be obtained by including the covariance in the formulation as
\begin{equation*}
    \hat{x}_{m|d} = \arg \min_{x_m} (Y-\Phi x_m)^\intercal \Omega(S_\mathrm Q)^{-1} (Y-\Phi x_m).
\end{equation*}
Since the process noise is state-dependent in our case, the state estimation problem cannot be approached directly. The matrix $\Omega(S_\mathrm Q)$ will be instead  estimated. For this purpose, introduce
the set $\hat S_{\mathrm Q} (x_m)$ of approximated process noise covariance matrices  as
\begin{equation*}
    \hat S_{\mathrm Q} (x_m)=\{Q(F^{k-m} x_m)\}_{k=0}^d.
\end{equation*}
A simplified version of the estimation problem can then be expressed as
\begin{equation} \label{eq:gmc}
    \hat{x}_{m|d} = \arg \min_{x_m} (Y-\Phi x_m)^\intercal \Omega(\hat S_{\mathrm Q}(x_m))^{-1} (Y-\Phi x_m).
\end{equation}
Since $\Omega(\hat S_{\mathrm Q}(x_m))$ depends on $x_m$, this problem is nonlinear. In this work, its solution is sought iteratively by applying Algorithm~\ref{alg:nls}.

\begin{algorithm}
\caption{Nonlinear least squares}\label{alg:nls}
\begin{algorithmic}
\State Solve $x = \arg \min\limits_{x_m} ||Y-\Phi x_m||^2$
\State let $s = \infty$, set $s_\mathrm{tol}$
\Repeat
    \State let $s_0=s$
    \State Calculate $\Omega(\hat Q(x))$
    \State Solve $\hat{x}=\arg \min\limits_{x_m} (Y-\Phi x_m)^\intercal \Omega(\hat Q(x))^{-1} (Y-\Phi x_m)$
    \State Let $s=(Y-\Phi \hat{x})^\intercal \Omega(\hat Q(x))^{-1} (Y-\Phi \hat{x})$
    \State Let $x=\hat x$
\Until{$|s_0-s|<s_\mathrm{tol}$}
\State Let $\hat{x}_{m|d}=x$
\end{algorithmic}
\end{algorithm}

\subsection{Numerical consideration} \label{sec:numCons}

For the parametrizations of $F$ that appear in this work, the observability matrix $\Phi$ that is utilized in solving the least squares problems of state estimation becomes numerically infeasible to calculate if $m$ is too large. The reason for this is that $F$ has eigenvalues that are significantly smaller than one in magnitude, so that repeated inversions result in very large elements in $\Phi$. To avoid this problem, the number of elements that was included in the optimization formulations was limited. For parametrizations, where one eigenvalue of $F$ is very close to zero, the solution was to remove the corresponding state through truncation, thus treating the state as identical zero.

The approximation $\hat x_k = F^{k-m} x_m$ in the nonlinear least squares formulation can also pose problems, when $k$ is significantly smaller than $m$. For this reason, a simple regularization was implemented, where $\hat x_k$ is set to zero whenever any element of $F^{k-m} x_m$ becomes negative.

\section{Experimental Results}\label{sec:experiments}

The three estimation algorithms introduced above were evaluated using two types of synthetic data. First,  linear model \eqref{eq:LTI} was used to generate the data, with the same Poisson-distributed state dependent noise sources as derived for the estimators. Then, the data were generated from stochastic simulations of the Markov chain model. In both cases, the models were simulated repeatedly over a time horizon of $42$ days ($d=42$), from identical initial conditions (distinct between the two cases) and with identical parameter values (identical between the two cases), that were randomly selected from the prior parameter distributions. The probability distributions of the state estimation errors for $m=30$ were  estimated by fitting  kernel distribution and compared to each other.

\subsection{Synthetic data from linear time-invariant model}\label{sec:synt_LTI}
The state estimation was performed with the three algorithms for $100$ realizations. The process noise covariance was calculated with the diagonal elements of $Q_0$ set to $0.1$, and $R_k=0.1$. Measurements for indices $k<19$ were neglected in the batch optimization approaches.

\begin{figure}
    \centering
    \includegraphics[width=8.5cm]{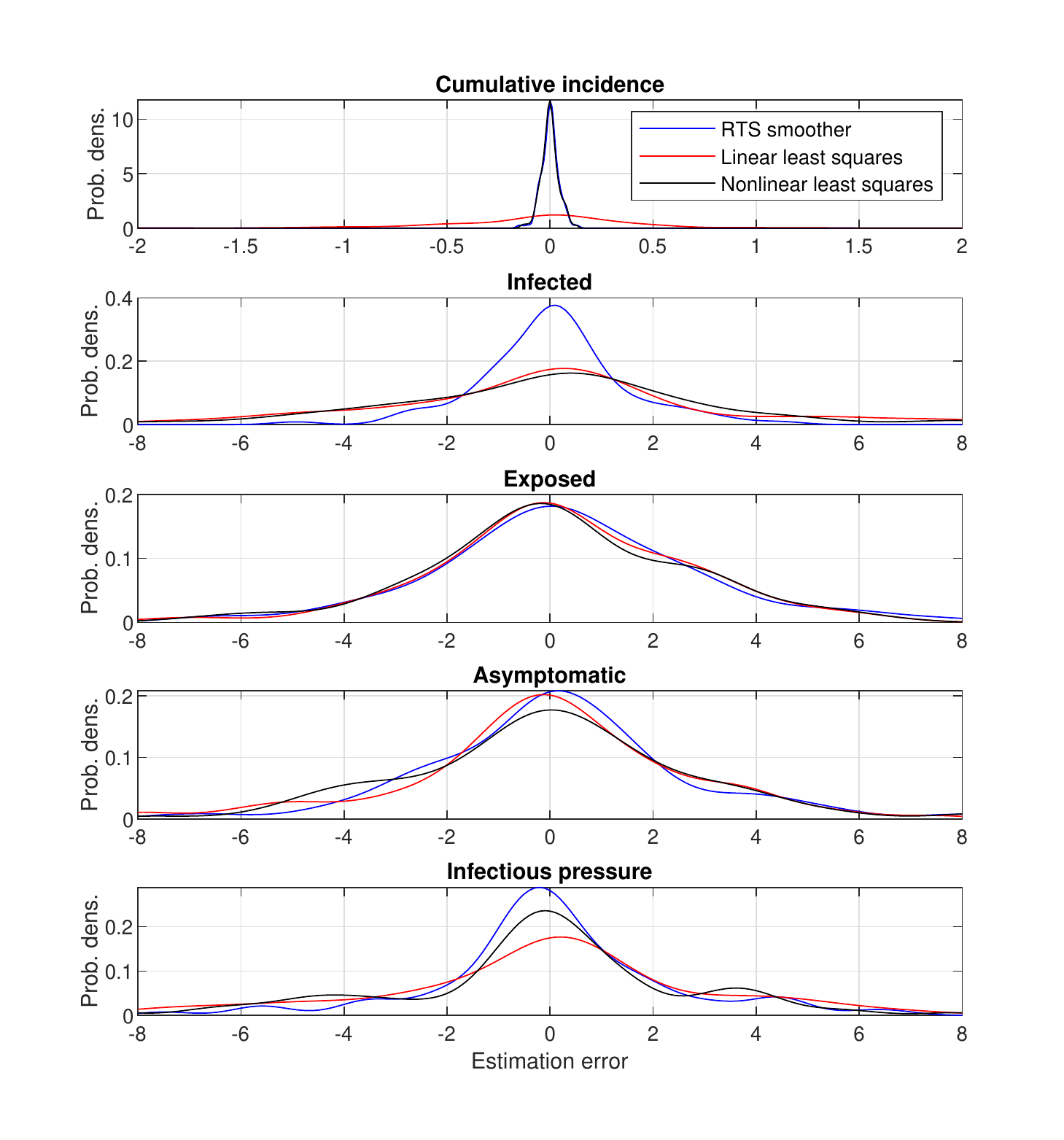}
    \caption{Estimation error probability density function for synthetic data from simulations of linear model \eqref{eq:LTI}.}
    \label{fig:linEst}
\end{figure}

The estimated distributions for all model states are shown in Fig.~\ref{fig:linEst}. The RTS smoother appears to perform the best, mostly through lower uncertainty in  the infected population estimate. The main difference between the linear and nonlinear least squares formulations is the significantly higher uncertainty in the cumulative incidence estimation for the linear method. This makes sense as the linear method does not exploit the low uncertainty of the measurement of this state, that is encoded in the covariance model.

\subsection{Synthetic data from the Markov chain model}\label{seq:syntMark}
In this case, $50$ realizations were generated and data from the three counties that were subject to spread of the disease in the highest number of realizations ($33$, $33$ and $31$ respectively) were analyzed. To capture the larger model discrepancy, the diagonal elements of $Q_0$ were set to $2$ and $R_k=0.5$. As above, indices $k<19$ were neglected in the batch optimizations.

\begin{figure}
    \centering
    \includegraphics[width=8.0cm]{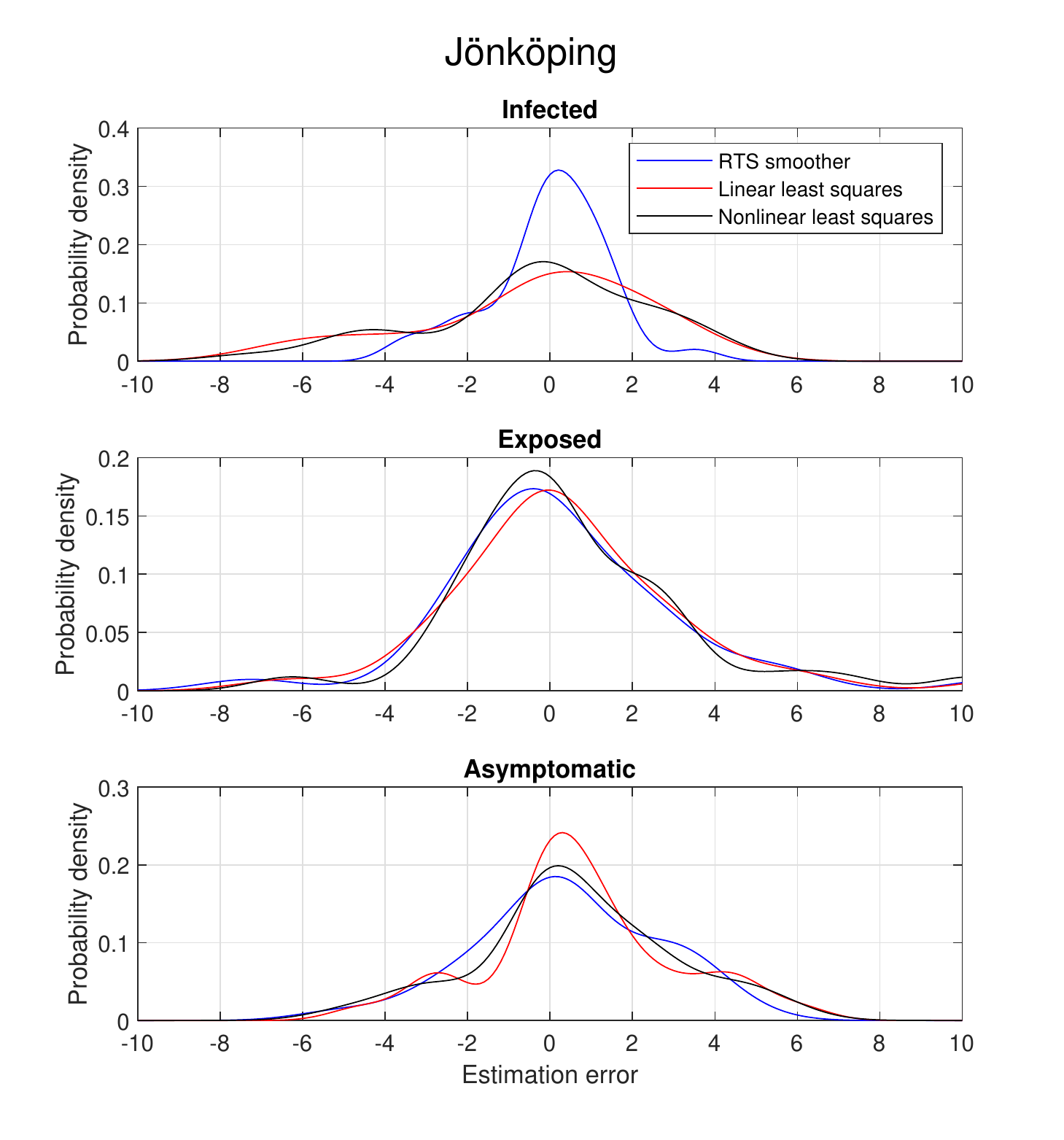}
    \caption{Estimation error probability density function for Jönköping county based on synthetic data from simulations of the Markov chain model.}
    \label{fig:jonkPDF}
\end{figure}

\begin{figure}
    \centering
    \includegraphics[width=8.0cm]{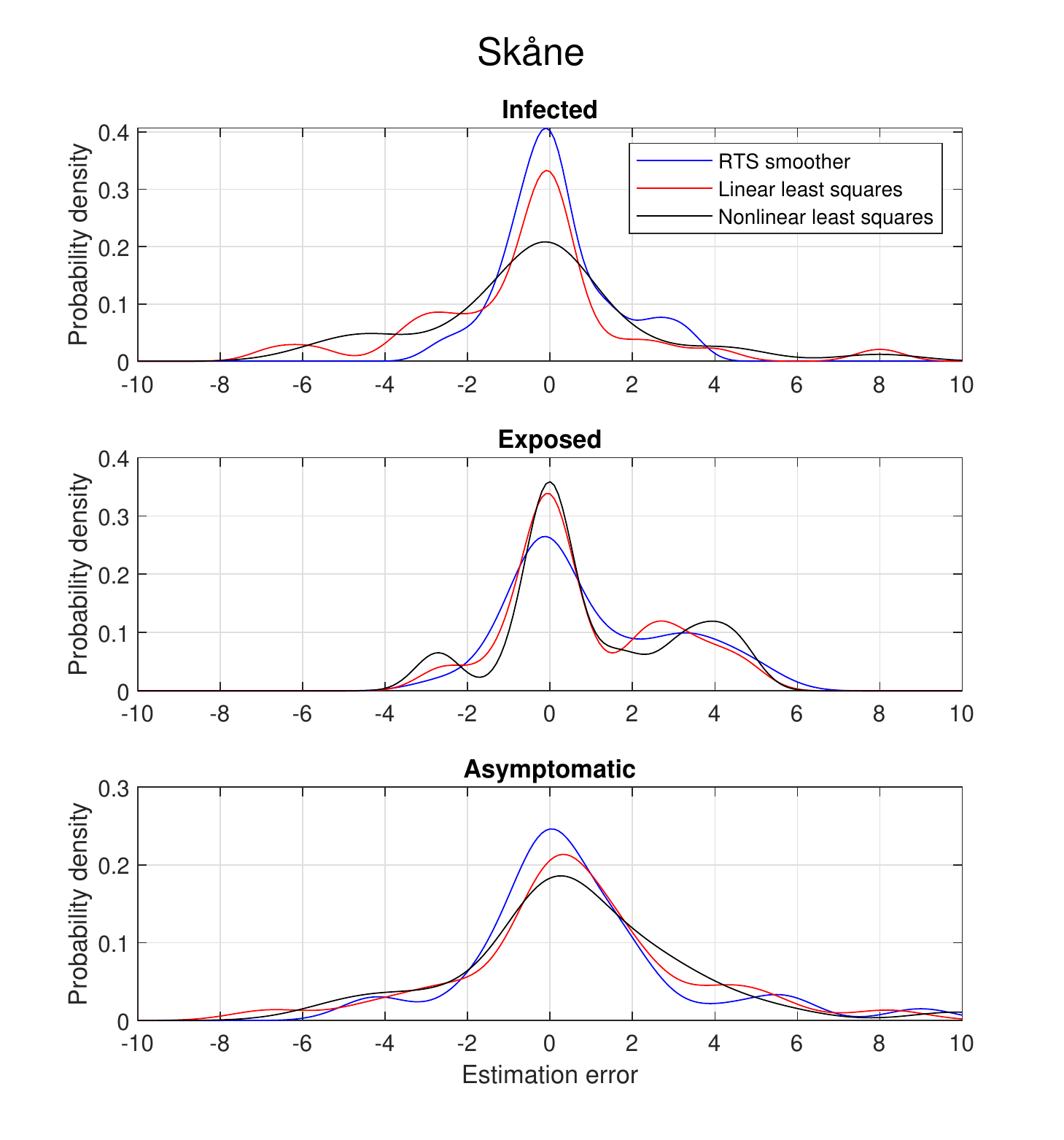}
    \caption{Estimation error probability density function for Skåne county based on synthetic data from simulations of the complete model.}
    \label{fig:skanePDF}
\end{figure}

\begin{figure}
    \centering
    \includegraphics[width=8.0cm]{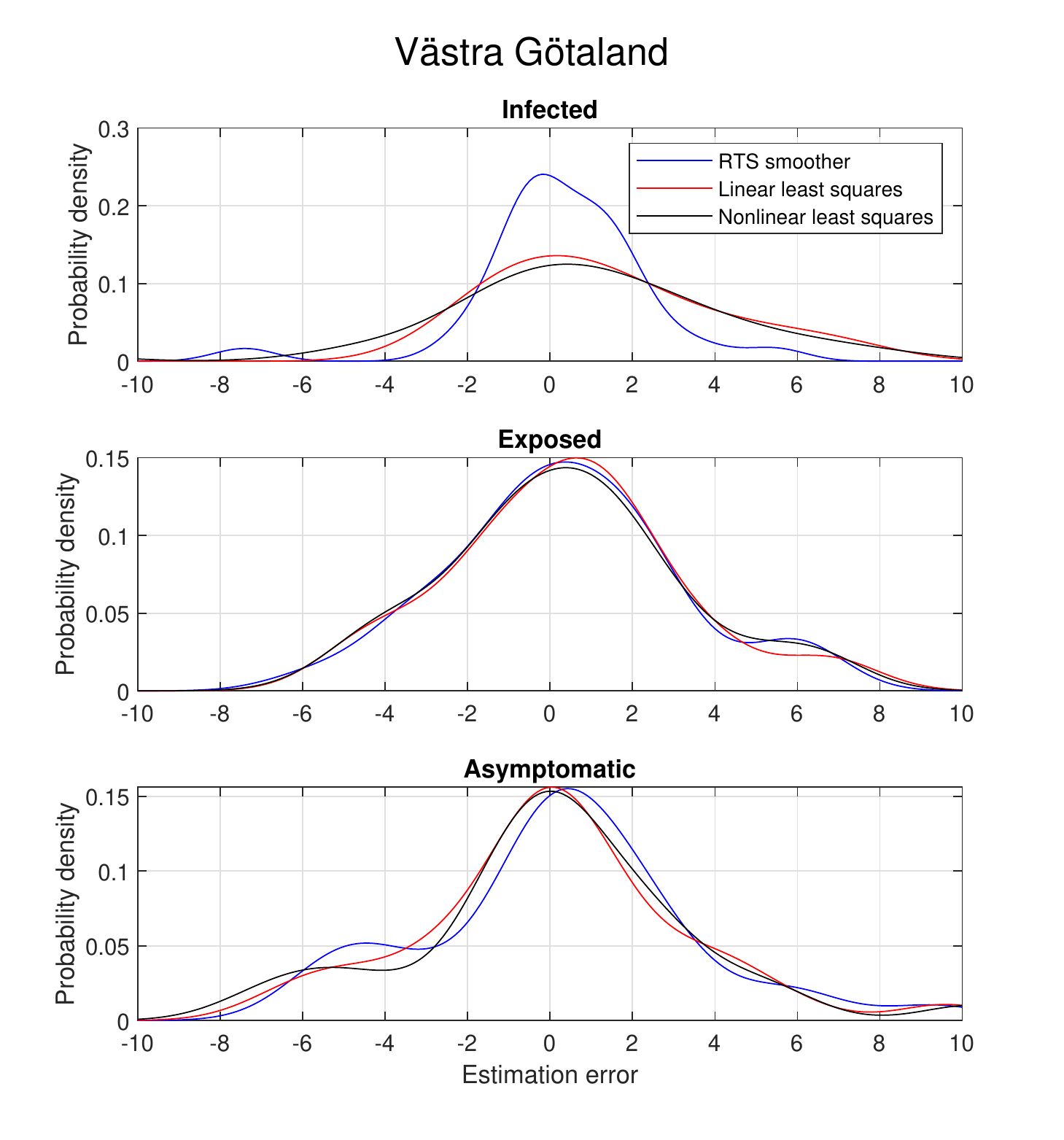}
    \caption{Estimation error probability density function for Västra Götaland county based on synthetic data from simulations of the complete model.}
    \label{fig:VGPDF}
\end{figure}

The estimated estimation error distributions for the states $I$, $E$ and $A$ in the three counties are shown in Fig.~\ref{fig:jonkPDF} -- Fig.~\ref{fig:VGPDF}. 

Similarly to the case considered in Section~\ref{sec:synt_LTI}, the RTS smoother is generally better at estimating the infected population. It is hard to draw conclusions apart from this from the plots, as the characteristics of the distributions vary between the counties.  

\subsection{Initialization of the Markov chain model with different estimates}
To investigate the effect of the initial estimation on the complete model, this model was simulated using estimated states as initial conditions. The estimated states from the three estimation methods for one realization of the simulation of the complete model were chosen. These are summarized in Table \ref{tab:ests}. The complete model was simulated $50$ times from each of the three sets of initial conditions, for $42$ days.

\begin{table}[ht]
    \centering
    \caption{State estimates in initalization experiment. Counties not included in the list had no estimated spread of the disease. Västra G. denotes Västra Götaland county and RTS, OLS and NLS denote state estimations using the RTS smoother, ordinary least squares and nonlinear least squares respectively.}\label{tab:ests}
    \begin{tabular}{c|ccc|ccc|ccc}
     & \multicolumn{3}{c}{RTS} & \multicolumn{3}{c}{OLS} & \multicolumn{3}{c}{NLS} \\
     County &$I$ &$E$ &$A$ & $I$ &$E$ &$A$ & $I$ &$E$ &$A$\\
     \hline
    Stockholm & 2&3&3&0&4&4&0&4&4 \\
    Skåne & 24&28&24&7&30&31&12&30&29 \\
    Västra G. & 24&39&30&36&40&28&29&40&31 \\
    \end{tabular}
    
    \label{tab:my_label}
\end{table}

\begin{figure}
    \centering
    \includegraphics[width=8.0cm]{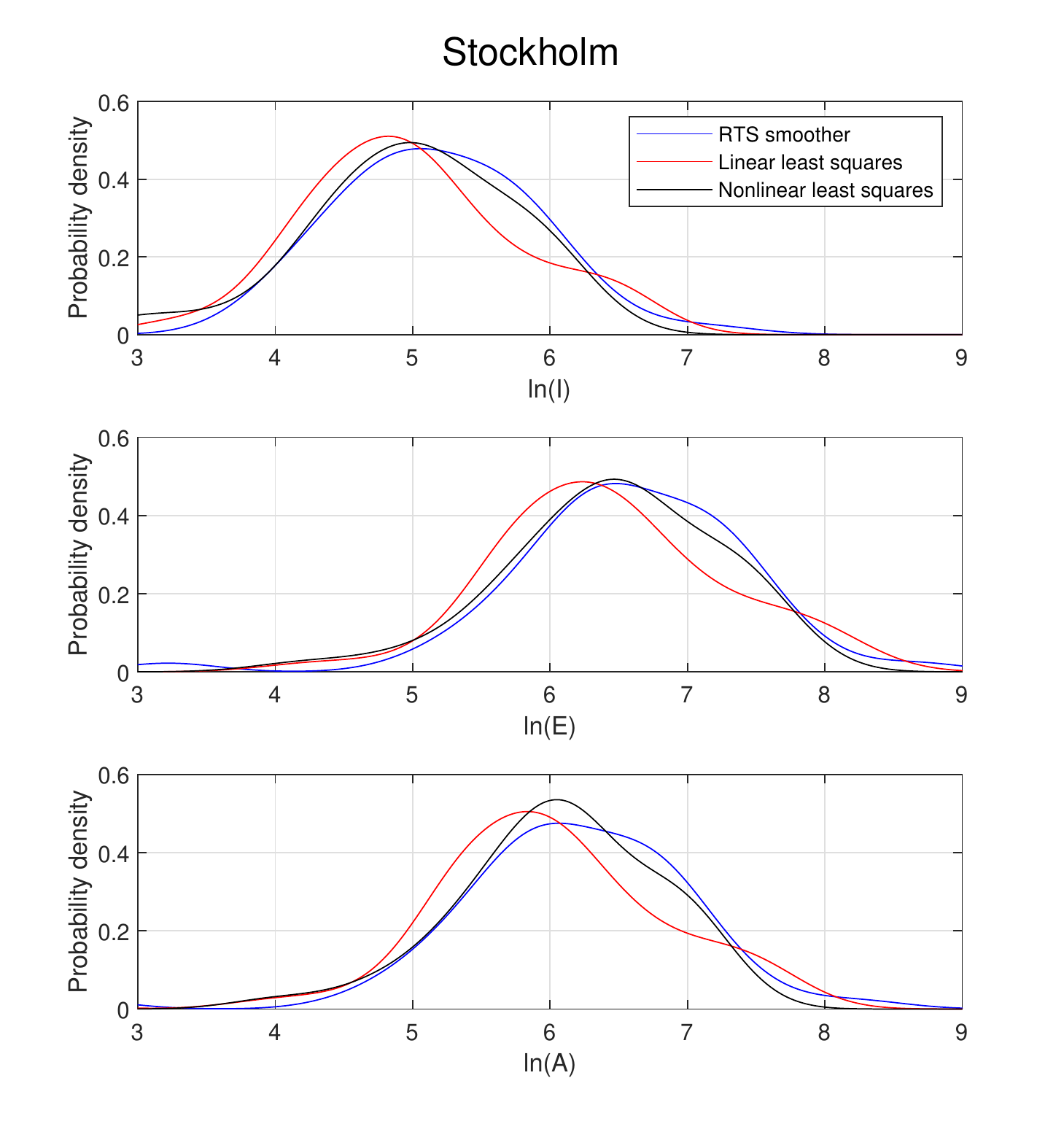}
    \caption{Probability distributions of model states for Stockholm county according to simulation from estimated initial conditions.}
    \label{fig:sthlmFinal}
\end{figure}

\begin{figure}
    \centering
    \includegraphics[width=8.0cm]{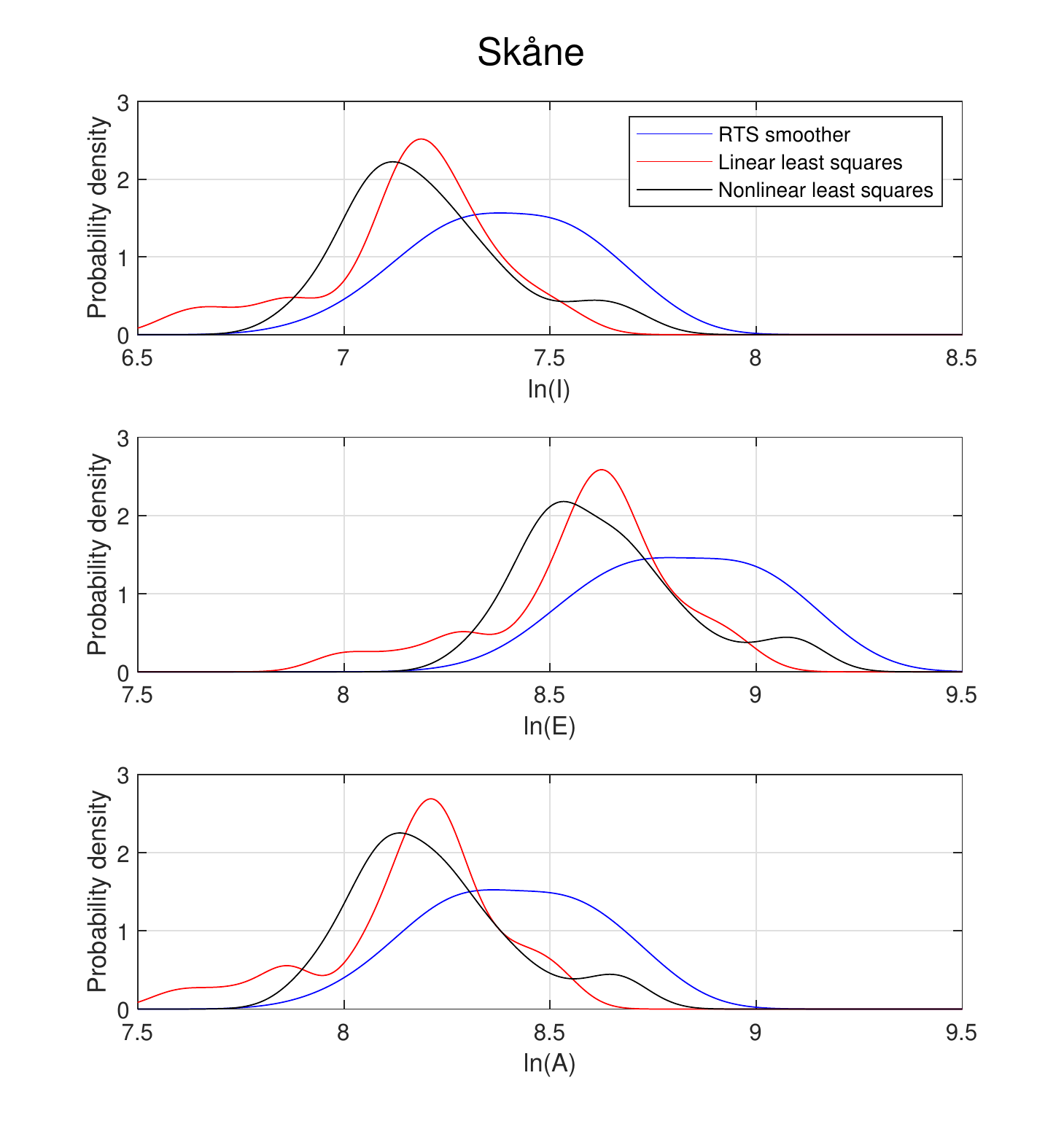}
    \caption{Probability distributions of model states for Skåne county according to simulation from estimated initial conditions.}
    \label{fig:skaneFinal}
\end{figure}

\begin{figure}
    \centering
    \includegraphics[width=8.0cm]{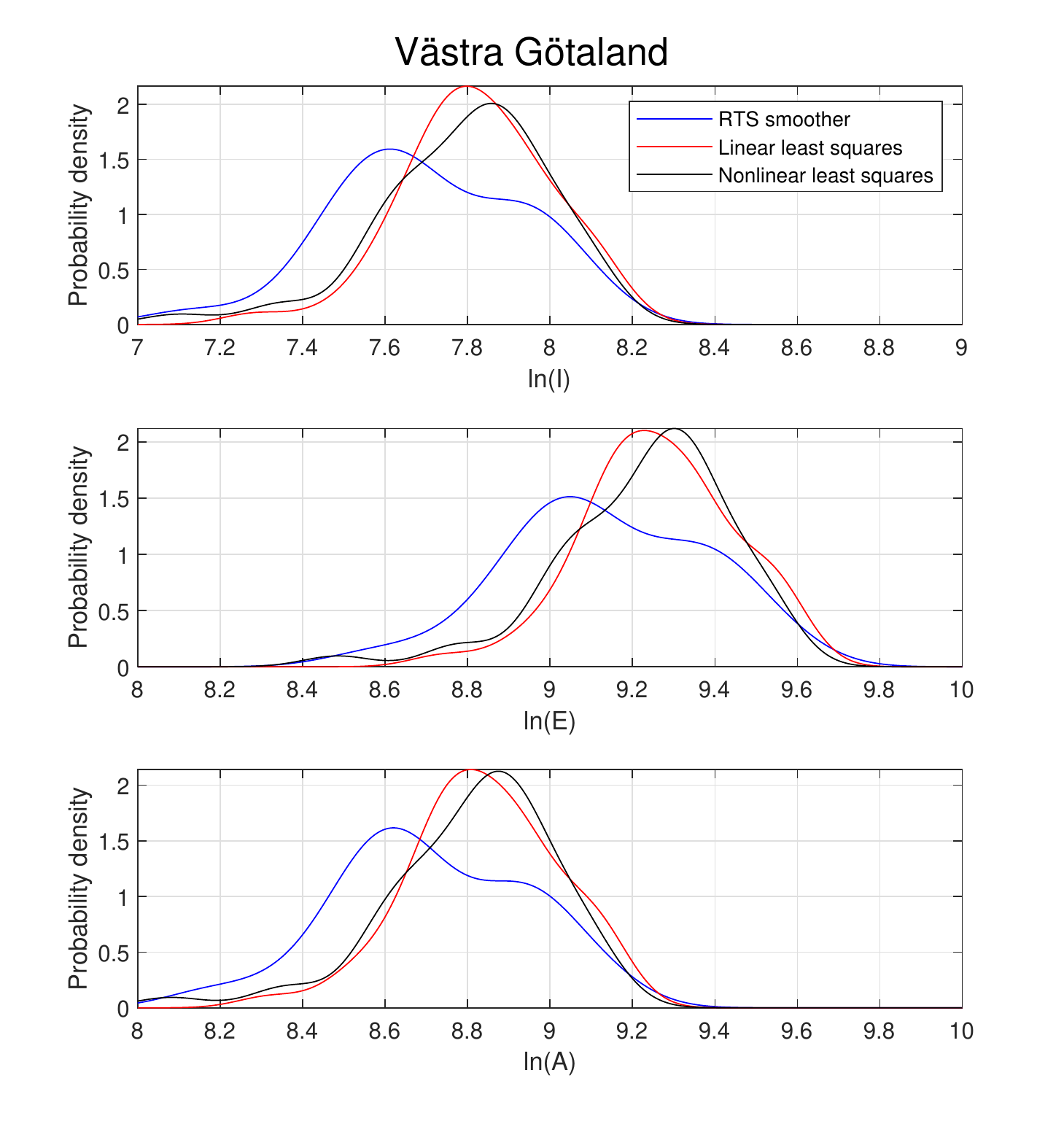}
    \caption{Probability distributions of model states for Västra Götaland county according to simulation from estimated initial conditions.}
    \label{fig:VGFinal}
\end{figure}

The probability distributions of the logarithm of the infected, exposed and asymptomatic populations, in the three counties listed in Table~\ref{tab:ests}, were then estimated using kernel distribution fitting. The results are depicted in Fig.~\ref{fig:sthlmFinal} -- Fig.~\ref{fig:VGFinal}.

The main conclusion that can be drawn from these results is that the variations between the considered estimation algorithms have limited effect on the states of the system in the end of the simulation, compared to the variations due the stochastic simulation. A greater variance in the states can be observed for the initial conditions generated by the RTS smoother compared to the other, but no general conclusion regarding the initialization methods can be drawn from this, as the results are based on a single estimation instance.

\section{Conclusion}
Three approaches to a fixed interval smoothing problem with the purpose of initialization of a larger epidemiological model have been compared; one based on the Rauch-Tung-Striebel smoother and two batch optimization methods. The non-Gaussian state-dependent noise in the model implies that standard approaches could not be used directly, instead covariance estimates were used in two of the methods. The results indicate that the smoother performs better than the other methods, despite the lack of theoretical justification of the method.

Simulations from estimated initial conditions indicate that the effect of minor estimation errors is limited compared to the variations inherit to the stochastic simulation of the Markov chain model. This suggests that computational complexity, robustness and ease of implementation might be of greater importance than high accuracy, when the initialization algorithm is chosen.


\bibliography{ifacconf}             

\end{document}